\documentclass{aastex}          
\usepackage{spr-astr-addons}    

\usepackage{color}

\begin{document}
%
\title{Acceleration of particles by black hole with
gravitomagnetic charge immersed in magnetic field}

\shorttitle{Acceleration of particles} \shortauthors{Abdujabbarov
et al.}


\author{A.A. Abdujabbarov\altaffilmark{1,2}} \email{ahmadjon@astrin.uz}  \and
\author{A.A. Tursunov\altaffilmark{1,2,3}}   \email{armantursun@gmail.com} \and
\author{B.J. Ahmedov\altaffilmark{1,2}} \email{ahmedov@astrin.uz}
\author{A. Kuvatov\altaffilmark{4}}

\altaffiltext{1}{Institute of Nuclear Physics,
        Ulughbek, Tashkent 100214, Uzbekistan}
\altaffiltext{2}{Ulugh Begh Astronomical Institute,
Astronomicheskaya 33, Tashkent 100052, Uzbekistan}
\altaffiltext{3}{National University of Uzbekistan,  Tashkent
100174, Uzbekistan} \altaffiltext{4}{Samarkand State University,
Samarkand 140104, Uzbekistan}

\begin{abstract}

The collision of test charged particles in the vicinity of an
event horizon of a non-rotating black hole with gravitomagnetic
charge immersed in external magnetic field has been studied. The
presence of the external magnetic field decreases the innermost
stable circular orbits (ISCO) radii of charged particles. The
opposite mechanism occurs when the gravitomagnetic charge of a
black hole is nonvanishing. For a collision of charged particle
moving at ISCO and the neutral particle falling from infinity the
maximal collision energy can be decreased by gravitomagnetic
charge in the presence of external asymptotically uniform magnetic
field.

\end{abstract}


\section{Introduction}\label{intro}

At present there is no any observational evidence for the
existence of gravitomagnetic monopole, i.e. so-called NUT
\citep{nut63} parameter or {\it magnetic mass}. Therefore study of
the motion of the test particles and particle acceleration
mechanisms in NUT spacetime may provide new tool for studying new
important general relativistic effects which are associated with
nondiagonal components of the metric tensor and have no Newtonian
analogues (See,
e.g.~\cite{nzlb97,lbnz98,nzdlb99,zonoz07,kkl08,kkhl10,ma08} where
solutions for electromagnetic waves and interferometry in
spacetime with NUT parameter have been studied.). Kerr-Taub-NUT
spacetime with Maxwell and dilation fields is recently
investigated by authors of the paper ~\cite{aliev08}. In our
preceding papers~\cite{mak08,aak08} we have studied the plasma
magnetosphere around a rotating, magnetized neutron star and
charged particle motion around compact objects immersed in
external magnetic field in the presence of the NUT parameter. The
Penrose process in the spacetime of rotating black hole with
nonvanishing gravitomagnetic charge has been considered in
\cite{shaymiev11}. The electromagnetic field of the relativistic
star with nonvanishing gravitomagnetic charge has been considered
by~\citet{aka12}.

Astrophysical processes which may produce high energy radiation
near a rotating black hole horizon attract more attention in
recent publications. The processes which are related to the effect
of
Penrose~\citep{pnr74} have been properly considered in
\citet{Piran:1975,Piran:1977dm,Piran:1977aa}. Recently
\citet{Banados:2009pr} (BSW) pointed
out that the collisions of
particles near extremely rotating
 black hole can produce
particles of high
center-of-mass energy. The results of
\citet{Banados:2009pr} have been commented by \cite{berti} where
authors concluded that astrophysical limitations on the maximal
spin, back-reaction
effects and sensitivity to the initial conditions impose severe
limits on the
likelihood of such accelerations. The acceleration
of particles, circular geodesics, accretion disk, and high-energy
collisions in the Janis-Newman-Winicour spacetime have been
considered~by~\citet{joshi1,joshi2}. The above mentioned
discussions forces us to study the particle acceleration in the
spacetime of a black hole with nonvanishing gravitomagnetic
charge.

Our aim in the present
paper is to examine an above
mentioned effect of particle acceleration in the presence of
gravitomagnetic charge for the case when the collision of
particles occurs in the vicinity of nonrotating black hole
embedded in magnetic field. It is very interesting to study the
electromagnetic fields and particle motion in NUT space with the
aim to get new tool for studying new important general
relativistic effects. Moreover this demonstration can be
interesting because of the existence of both theoretical
and experimental evidences that a magnetic
field must be
present in the
vicinity of black
holes. Note, that hereafter we use the weak magnetic field
approximation in such sense that the
energy-momentum of this field
does
not change the background
geometry of black hole. For a
black hole with
mass $M$ this
condition means that the strength
of magnetic
field satisfies to the following
condition~\citep{psgn11,frolov}
\begin{equation}
\label{BBB}
B\ll B_{max}={c^4\over G^{3/2}
M_{\odot}}\left(\frac{M_{\odot}}{M}\right)\sim
10^{19}{M_{\odot}\over M}\mbox{Gauss}\, .
\end{equation}
Following to~\citet{frolov}, let us call these black
holes as {\em weakly magnetized}. One can say that this condition
is quite general and satisfies for both stellar
mass and
supermassive
black holes.

The paper is organized as follows. Section~II is devoted to study
of  electromagnetic field and charged particle motion in the
magnetized black holes with NUT parameter, with the main focus on
the properties of their ISCOs. Particles collisions in the
vicinity of a weakly magnetized black hole with nonvanishing
gravitomagnetic charge have been studied in Section~III. The
concluding remarks and discussions are presented in Section~IV.

Throughout the paper, we use a space-like signature $(-,+,+,+)$
and a system of units in which $G = 1 = c$ (However, for those
expressions with an astrophysical application we have written the
speed of light explicitly.). Greek indices are taken to run from 0
to 3 and Latin indices from 1 to 3; covariant derivatives are
denoted with a semi-colon and partial derivatives with a comma.

\section{Charged particle motion around black hole
with nonvanishing NUT charge}

\label{potential}

Here we will consider  a charged particle motion in the vicinity
of a black hole of mass $M$ with gravitomagnetic charge in the
presence of an external
axisymmetric and uniform
at the spatial
infinity magnetic field. The appropriate spacetime metric has the
following form~\citep[see][]{Dadhich02,bini03}:
\begin{eqnarray}
\label{metric}
ds^2&=&-\frac{\Delta}{\Sigma}\ dt^2+4\frac{\Delta}{\Sigma}\
l\cos\theta dt d\varphi+\frac{1}{\Sigma}\ (\Sigma^2\sin^2\theta-{}
\nonumber\\
&& {}-4\Delta l^2\cos^2\theta)d\varphi^2+ \frac{\Sigma}{\Delta}\
dr^2+\Sigma d\theta^2\ ,
\end{eqnarray}
where parameters $\Sigma$ and $\Delta$ are defined as
$$
\Sigma=r^2+l^2,\qquad\Delta=r^2-2 M r-l^2\ ,
$$
where $l$ is the gravitomagnetic monopole momentum. Here we will
exploit the existence in this spacetime of a timelike Killing
vector $\xi^{\alpha}_{(t)}=\partial x^\alpha /\partial t$ and
spacelike one $\xi^{\alpha}_{(\varphi)}=\partial x^\alpha
/\partial \varphi$ being responsible for stationarity and axial
symmetry of geometry, such that they satisfy the Killing equations
\begin{equation}
\xi_{\alpha;\beta}+\xi_{\beta;\alpha}=0
\end{equation}
which gives a right to write the solution of vacuum Maxwell
equations $\Box A^{\mu}=0$ for the vector potential $A_{\mu}$ of
the electromagnetic field in the Lorentz gauge in the simple form
\begin{equation}
A^{\alpha}=C_{1}\xi^{\alpha}_{(t)}+C_{2}\xi^{\alpha}_{(\varphi)}.
\end{equation}
The constant $C_{2} = B/2$, where gravitational source is immersed
in the uniform magnetic field $\mathbf{B}$ being parallel to its
axis of rotation. The value of the remaining constant $C_1$ can be
easily calculated from the asymptotic properties of spacetime
(\ref{metric}) at the infinity.
Indeed in order to find the remaining constant one can use the
electrical neutrality of the black hole
\begin{eqnarray}
\label{flux} 4\pi Q&=&\frac{1}{2}\oint
F^{\alpha\beta}{_*dS}_{\alpha\beta}= C_1 \oint
\Gamma^\alpha_{\beta\gamma}\tau_\alpha m^\beta
\xi^\gamma_{(t)}(\tau k)dS \nonumber\\
&&
+\frac{B}{2}\oint \Gamma^\alpha_{\beta\gamma}\tau_\alpha m^\beta
\xi^\gamma_{(\phi)}(\tau k)dS = 0\ ,
\end{eqnarray}
evaluating the value of the integral through the spherical surface
at the asymptotic infinity. Here the equality
$\xi_{\beta;\alpha}=-\xi_{\alpha;\beta}=
-\Gamma^\gamma_{\alpha\beta}\xi_\gamma$ following from the Killing
equation was used, and element of an arbitrary 2-surface
$dS^{\alpha\beta}$ is represented in the form~\citep[see, e.g.\
][]{ar03}
\begin{equation}
\label{surface} dS^{\alpha\beta}=-\tau^\alpha\wedge m^\beta (\tau
k)dS+\eta^{\alpha\beta\mu\nu}\tau_\mu n_\nu\sqrt{1+(\tau k)^2}dS\
,
\end{equation}
and the following couples
\begin{eqnarray}
& m_\alpha =&\frac{\eta_{\lambda\alpha\mu\nu}\tau^\lambda n^\mu
k^\nu}
{\sqrt{1+(\tau k)^2}}, \nonumber\\
&n_\alpha =&\frac{\eta_{\lambda\alpha\mu\nu} \tau^\lambda k^\mu
m^\nu}{\sqrt{1+(\tau k)^2}}, \quad \nonumber\\
&k^\alpha=&-(\tau k)\tau^\alpha+ \sqrt{1+(\tau
k)^2}\eta^{\mu\alpha\rho\nu}\tau_\mu m_\rho n_\nu \nonumber
\end{eqnarray}
are established between the triple $\{{\mathbf k},{\mathbf
m},{\mathbf n}\}$ of vectors, $n^\alpha$ is normal to 2-surface,
space-like vector $m^\alpha$ belongs to the given 2-surface and is
orthogonal to the four-velocity of observer $\tau^\alpha$, a unit
spacelike four-vector $k^\alpha$ belongs to the surface and is
orthogonal to $m^\alpha$, $dS$ is invariant element of surface,
$\wedge$ denotes the wedge product, $_*$ is for the dual element,
$\eta_{\alpha\beta\gamma\delta}$ is the pseudo-tensorial
expression for the Levi-Civita symbol $\epsilon_{\alpha \beta
\gamma \delta}$.

Then one can insert $\tau_0=-(1-M/r)$, $m^1=(1-M/r)$, and
asymptotic values for the Christoffel symbols
$\Gamma^0_{10}=M/r^2$ and $\Gamma^0_{13}=-l(1-2M/r)\cos\theta/r$
in the flux expression (\ref{flux}) and get the value of constant
$C_1=0$. Parameter $l$ does not affect on constant $C_1$ because
the integral $\int_0^{\pi}\cos\theta\sin\theta d\theta=0$
vanishes.

Thus the 4-vector potential $A_{\mu}$ of the electromagnetic field
will take the following form
\begin{eqnarray}
&& A_{0}=-\frac{\Delta}{\Sigma} B l \cos{\theta} \\
&& A_{3}=\frac{1}{2} \Sigma^2 \sin^2{\theta}-2 \Delta
l^2\cos^2{\theta}
\end{eqnarray}

The orthonormal components of the electromagnetic fields measured
by zero angular momentum observers (ZAMO) with four velocity
components
\begin{eqnarray}
&&
\hspace{-0.5cm}(\tau^\alpha)_{\textrm{ZAMO}}\equiv\left(\sqrt{\frac{{\cal
R}}{\Delta\Sigma \sin^2\theta}}, 0, 0, \frac{2\Delta l
\cos\theta}{\sqrt{\Delta\Sigma {\cal R}\sin^2\theta}}\right)\ \ \
\\
&&\hspace{-0.5cm} (\tau_{\alpha})_{
\textrm{ZAMO}}\equiv\left(\sqrt{\frac{\Delta\Sigma\sin^2{\theta}}{{\cal
R}}}, 0, 0, 0\right) \end{eqnarray}
are given by expressions
\begin{eqnarray}
\label{ehr}E^{\hat{r}}&=&-\frac{B r  l}{\sqrt{{\cal R}}}
\left(1-\frac{M}{r}\right) \sin 2\theta\ ,
\\
E^{\hat{\theta}}&=&\frac{B l}{\Sigma^2}\sqrt{\frac{\Delta}{{\cal
R}}} \nonumber
\\&&\times \left[\Sigma^2+\left(\Sigma^2-2\Delta l^2 \cos\theta\right)
\label{eht} \frac{2\cos\theta}{\sin^2\theta}\right]\sin^2\theta\ ,
\\
\label{bhr}B^{\hat{r}}&=&\frac{B\tan\theta}{\Sigma\sqrt{{\cal R}}}
\left({\cal R}-\Sigma^2 \right)\ ,
\\
B^{\hat{\theta}}&=&\frac{B r}{\Sigma^2} \sqrt{\frac{\Delta}{{\cal
R}}}
 \cos^2\theta \nonumber\\&&\times \left\{\left[\Delta-\Sigma\left( 1-\frac{M}{r}
\right)\right] 4 l^2 +\Sigma^2 \tan^2\theta\right\} \ ,\label{bht}
\end{eqnarray}
where the following notation has been used:
$$ {\cal
R}=\Sigma^2 \sin^2{\theta}-4\Delta l^2 \cos^2{\theta}. $$

Astrophysically it is interesting to know the limiting cases of
expressions (\ref{ehr}) –- (\ref{bht}), for example in either
linear or quadratic approximation ${\mathcal O}(a^{2}/r^{2},
l^{2}/r^{2})$ in order to give physical interpretation of possible
physical processes near the slowly rotating relativistic compact
objects, where they take the following form:
\begin{eqnarray}
\label{eehr}E^{\hat{r}}&=&\frac{2Bl\cos{\theta}}{r}\left(1
-\frac{3M}{r}\right) \ ,
\\
\label{eeht}E^{\hat{\theta}}&=&\frac{Bl\sin\theta}{r}\left(1
+\frac{2\cos\theta}{\sin^2\theta}\right)\ ,
\\
\label{bbhr}B^{\hat{r}}& = & B\cos\theta \left[1+2\frac{l^2}{r^2}
\frac{1}{\sin^2{\theta}}\right]\ ,
\\
B^{\hat{\theta}} &= &B \sin\theta \bigg[1-\frac{M}{r}
+\frac{1}{16r^2 \sin^2\theta}\nonumber\\&& \times
\Big(4l^2-4M^2+4(7l^2+M^2)\cos2\theta\Big)\bigg]\ .\label{bbht}
\end{eqnarray}

In the limit of flat spacetime, i.e. for $M/r\rightarrow0$ and
$l^2/r^2\rightarrow0$, expressions (\ref{eehr})-(\ref{bbht}) give
\begin{equation}
\label{eflimit}E^{\hat{r}}=E^{\hat{\theta}}\rightarrow0\ ,
\end{equation}
\begin{equation}
\label{bflimit}B^{\hat{r}}\rightarrow B\cos\theta, \qquad
B^{\hat{\theta}}\rightarrow B\sin\theta\ .
\end{equation}
As expected, expressions (\ref{eflimit})-(\ref{bflimit}) coincide
with the solutions for the homogeneous magnetic field in Newtonian
spacetime.

The dynamical
equation for a
charged
particle motion
can be
written as
\begin{equation}
m{du^{\mu}\over d\tau}=qF^{\mu}_{\,\,\,\nu}\,u^{\nu}\,,\label{1}
\end{equation}
where $\tau$ is the proper
 time, $u^{\mu}$ is the 4-velocity of a
charged particle, $u^{\mu}u_{\mu}=-1$, $q$
and $m$ are its charge
and mass,
respectively. $F_{\alpha\beta} =
A_{\beta,\alpha}-A_{\alpha,\beta}$ is the antisymmetric tensor of
the electromagnetic field, which has the following four
independent components
\begin{eqnarray}
&& F_{01}=\frac{B}{\Sigma^2} 2 l[(r-M)-\Delta r] \cos\theta\ ,
\nonumber\\ && F_{02}=\frac{B}{\Sigma}\Delta l \sin\theta\ ,
\nonumber\\ && F_{13}=-\frac{B}{\Sigma^2} 8l^2[(r-M)\Sigma-\Delta
r]\cos^2{\theta}+B r \sin^2{\theta}\ ,\nonumber\\
&&
F_{23}=\frac{B}{\Sigma} (4l^2 \Delta+\frac{1}{2}\Sigma^2)
\sin{2\theta}\nonumber\ .
\end{eqnarray}
While considering the charged particle motion around black hole
immersed in the magnetic field it is easy to use two conserved
quantities associated with the Killing
vectors: the energy ${\cal
E}>0$ and
the generalized
azimuthal angular
momentum ${\cal L}
\in(-\infty,+\infty)$:
\begin{eqnarray} {\cal E} & \hspace{-.15cm} \equiv
-\xi^{\mu}_{(t)}P_{\mu}=& \hspace{-.35cm} \frac{m \Delta}{\Sigma}
(\frac{dt}{d\tau}+4l\cos\theta
\frac{d \varphi}{d\tau}+\frac{q}{m}Bl\cos{\theta}), \ \ \\
{\cal L}& \hspace{-.15cm} \equiv  \xi^{\mu}_{(\phi)}P_{\mu}=
&\hspace{-.35cm}- 4ml \frac{\Delta}{\Sigma} \cos{\theta}
\frac{dt}{d\tau}+(\Sigma\sin^2{\theta}-{}
\nonumber\\
&&  {\hspace{-.5cm}}-4 l^2 \frac{\Delta}{\Sigma} \cos^2{\theta})
\left(m\frac{d\varphi}{d\tau}+\frac{q B}{2}\right).
\end{eqnarray}
Here
$P_{\mu}=m\,u_{\mu}+q A_{\mu}$ is the
generalized
4-momentum of a charged
particle. It was first shown by \citet{Zim89} for spherical
symmetric case (NUT spacetime) and later by \citet{bini03} for
axial symmetric case (Kerr-Taub-NUT spacetime) that the orbits of
the test particles are confined to a cone with the opening angle
$\theta$ given by $\cos\theta = 2{\cal E}l/{\cal L}$. It also
follows that in this case the equations of motion on the cone
depend on $l$ only via $l^2$ \citep{bini03,aak08}. The main point
is that the small value for the upper limit for gravitomagnetic
moment has been obtained by comparing theoretical results with
experimental data as (i) $l\leq 10^{-24}$ from the gravitational
microlensing \citep{habibi04}, (ii) $l\leq 1.5\cdot10^{-18}$ from
the interferometry experiments on ultra-cold atoms \citep{ma08},
(iii) and similar limit has been obtained from the experiments on
Mach-Zehnder interferometer \citep{kkl08}. Due to the smallness of
the gravitomagnetic charge let us consider the motion in the
quasi-equatorial plane when the motion in $\theta$ direction
changes as $\theta=\pi/2+\delta\theta(t)$, where $\delta\theta(t)$
is the term of first order in $l$, then it is easy to expand the
trigonometric functions as $\sin\theta=1-\delta\theta^2(t)/2+{\cal
O}(\delta\theta^4(t))$ and $\cos \theta=\delta\theta(t)-{\cal
O}(\delta\theta^3(t))$. Neglecting the small terms ${\cal
O}(\delta\theta^2(t))$, one can easily obtain the geodesic
equation in the following form
\begin{eqnarray}
\label{teq}\frac{dt}{d\tau}&=& \frac{{\cal E}}{m} \frac{\Sigma}{\Delta}\ ,\\
\label{pheq}\frac{d\varphi}{d\tau}&=&\frac{{\cal
L}}{m\Sigma}-\frac{q B}{2m}\ ,\label{thetaeq}
\\
\label{req}\frac{d r}{d\tau}&=& \left(\frac{{\cal E}}{m}-U\right)\
,\end{eqnarray}
where $U$ denotes the effective potential as
\begin{equation}
\label{22} U=\frac{\Delta}{\Sigma}\left(1+\Sigma\chi^2\right)\ ,
\end{equation}
and
\begin{equation}
\chi=\frac{{\cal L}}{m\Sigma}-\frac{q B}{2m}\ .
\end{equation}

\begin{figure*}[htb]
\begin{center}
a) \includegraphics[width=8cm]{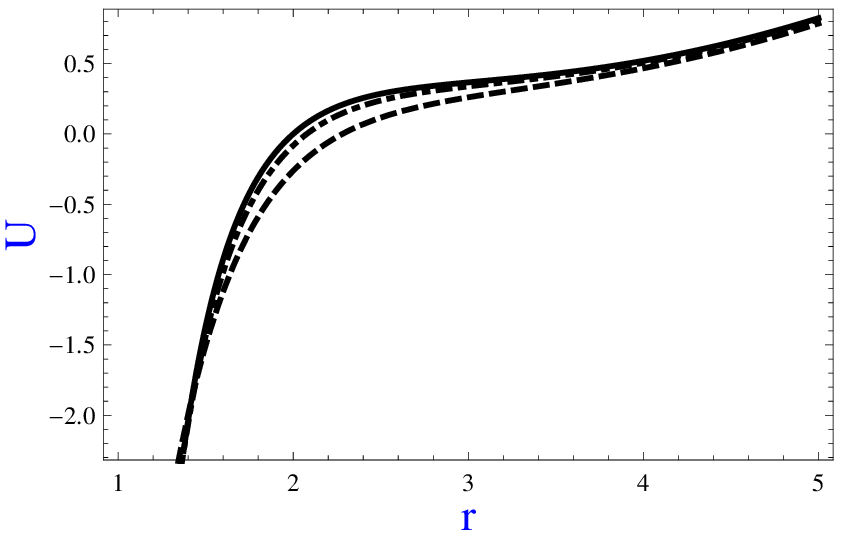} b)
\includegraphics[width=8cm]{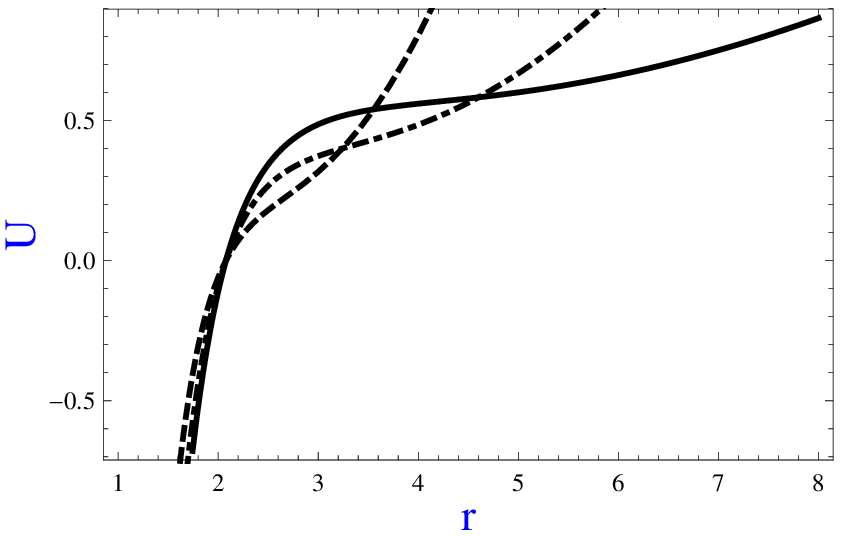}
\caption{\label{effpot} The radial dependence of the effective
potential of radial motion of the charged particle for the
different values of the gravitomagnetic charge (a): $l/M=0$ (solid
line), $l/M=0.4$ (dot-dashed line), $l/M=0.8$ (dashed line) and
for the different values of the dimensionless magnetic parameter
$b=qBM/m$ (b): $b=0.2$ (solid line), $b=0.4$ (dot-dashed line),
$b=0.8$ (dashed line). }\label{fig_0}
\end{center}
\end{figure*}

{In the expressions (\ref{teq})--(\ref{req}) the terms being
proportional to the second and higher orders of the small
parameter  $\delta\theta$ are neglected. In the Fig.~\ref{effpot}
the radial dependence of the effective potential of the radial
motion of the charged particle is presented for the different
values of dimensionless gravitomagnetic parameter $l/M$ and
magnetic parameter $b=qBM/m$. }

{From the Fig.~\ref{effpot} one can obtain the behavior of the
charged particle motion in the presence of both the
gravitomagnetic charge and magnetic field. In the presence of the
gravitomagnetic charge the minimum of the effective potential
shifts towards the observer at the infinity which means that the
orbits of the charged particles may become unstable. The minimum
value of the radius of the stable circular orbits increases. The
influence of the magnetic field has the opposite effect: the
presence of the magnetic field decreases the minimum value of the
circular orbits radius and charged particles may come much closer
to the central object.}

\begin{figure*}[htb]
\begin{center}
a) \includegraphics[width=8cm]{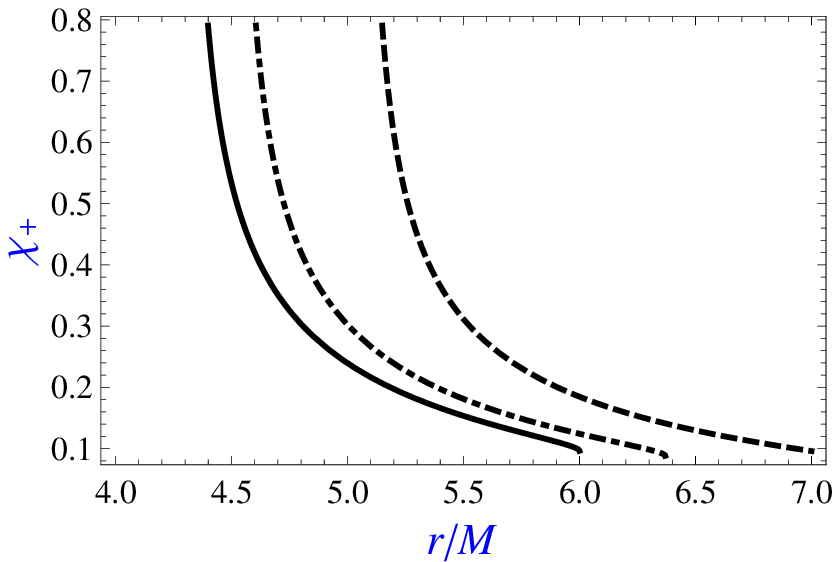} b)
\includegraphics[width=8cm]{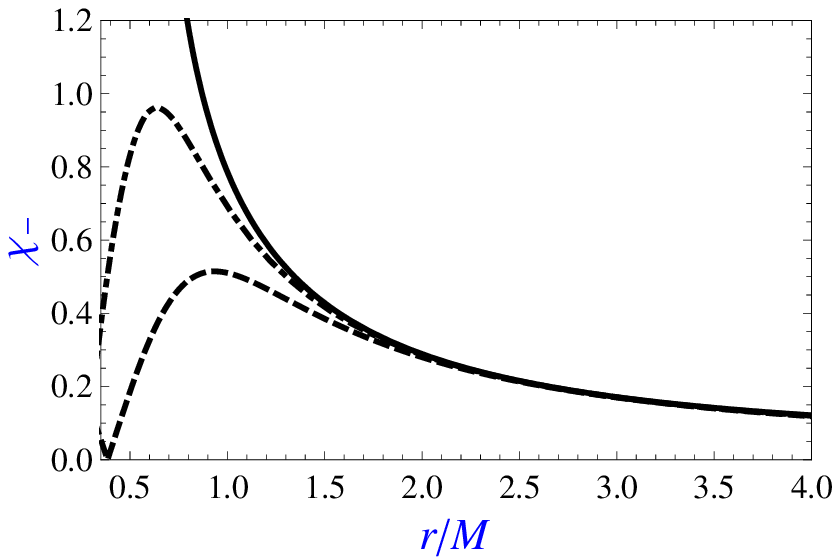}
\caption{\label{chi}$\chi_{+}$ (a) and $\chi_{-}$ (b) as a
function of the ISCO radius for different values of the
gravitomagnetic charge: $l/M=0$ (solid line), $l/M=0.4$
(dot-dashed line), $l/M=0.8$ (dashed line). }\label{fig_0}
\end{center}
\end{figure*}

In order to  study innermost stable circular orbits (ISCO)  we use
its first and second derivatives of the $U$ and equalize them to
zero:
\begin{eqnarray}
U'&=-&\frac{2\Delta r (1+2{\cal L}\chi)}{\Sigma^2}\nonumber\\&&
+\frac{1}{r} \left(1+\frac{\Delta}{\Sigma}\right)
\left(1+\Sigma\chi^2\right)=0
\label{ur}\\
U''&=& \frac{8\Delta L^2 r^2}{\Sigma^4}+\frac{2}{\Sigma}
\left(\frac{4\Delta r^2}{\Sigma^2}-\frac{3\Delta}{\Sigma}-2\right)
\nonumber\\
&& \times (1+2{\cal L} \chi)+{} +\frac{2}{ \Sigma} \left(1+
\Sigma\chi^2 \right)=0\label{urr}
\end{eqnarray}
Now we have two equations with three unknown quantities ${\cal
L}$, $r$ and $\chi$. Solving the equation (\ref{ur}) we derive
${\cal L}$ in terms $r$ and $\chi$
\begin{equation}
{\cal L}=\frac{-2 \Delta r^2+\Delta\Sigma+\Sigma^2+\Delta\Sigma^2
\chi^2+\Sigma^3 \chi^2}{4\Delta r^2 \chi}\ .
\end{equation}
Substituting this equation into (\ref{urr}) one can easily obtain
the equation  expressing  $\chi$ in terms of $r$
\begin{equation}
\chi_{\pm}=\pm\left[\frac{K}{\Sigma
A}\left(1\pm\sqrt{1-\frac{A\Sigma}{K^2}\left(\Delta +
\Sigma-\frac{2 r^2}{\Sigma}\Delta \right)^2
}\right)\right]^{\frac{1}{2}},
\end{equation}
 where
\begin{eqnarray}
&& A=8\Delta^2 r^2-5\Delta^2 \Sigma+12\Delta r^2
\Sigma-8\Delta\Sigma^2-3\Sigma^3, \nonumber\\
&& K=-2\Delta^2 r^2+2\Delta^2 \Sigma-4\Delta r^2
\Sigma+3\Delta\Sigma^2+\Sigma^3. \nonumber
\end{eqnarray}
{The signs $\pm$ correspond to co-rotating and counter-rotating
particle orbits, respectively.} The dependence of the $\chi_{\pm}$
from the ISCO radius are shown in Fig. \ref{chi}. The shift of the
shape of $\chi_{+}$ to the right direction with increasing the
gravitomagnetic charge corresponds to increasing of ISCO radius in
the presence of NUT charge.

{From the Fig. \ref{chi} one can see that $\chi_{-}$ function
steadily increase when particle approach to the black hole horizon
and go to infinity in the case when gravitomagnetic charge
vanishes. In the case of nonvanishing NUT-parameter, the function
$\chi_{-}$ falls in the point of singularity. But by reason of
that jumps are located inside the horizon, the effects that take
place there are not observable relatively to detached observer and
cannot be interpreted as some full physical theory. In other hand
the spacetime of nonrotating black hole allows us to find analytic
continuations of our theories inside the black hole horizon up to
a point of singularity by excepting it.}

We concentrate on a circular
motion of the charged
particle in the presence of NUT charge. At ISCO radius the
effective
potential has
minimum. The 4-momentum of a test charged particle at the
circular orbit
of radius
$r$ is (\cite{frolov})
\begin{eqnarray}
p^{\mu}&=&m \gamma (e_{(t)}^{\mu}+v e_{(\phi)}^{\mu})\, ,\label{pp}\\
e_{(t)}^{\mu}&=&(\Sigma/\Delta)^{1/2}\xi_{(t)}^{\mu}=(\Sigma/\Delta)^{1/2}\delta_t^{\mu}\, ,\\
e_{(\phi)}^{\mu}&=&\Sigma^{-1/2}\xi_{(\phi)}^{\mu}=\Sigma^{-1/2}\delta_{\phi}^{\mu}\,
.
\end{eqnarray}
%
Here $v$ is a
velocity of a charged
particle with
respect to a
rest frame. $v$ can be both
positive and
negative, and $\gamma$ is
the Lorentz
gamma factor, which is always positive. Using the condition of the
normalization for the momentum
 $\mathbf{p}^2=-m^2$ one
has $\gamma=(1-v^2)^{-1/2}$. For the positive charge $q>0$ the
Lorentz force
acting
on a charged particle
with
$v>0$ is {\it repulsive}, i.e. the force is directed
outwards
the black hole, while for
$v<0$ the Lorenz force is
{\it attractive}.

Following to \cite{frolov}, one can use an expression
$d\phi/d\tau=v\gamma/r$ and (\ref{thetaeq}) with $\theta=\pi/2$.
This implyies
\begin{equation}
\frac{v\gamma}{r}=\chi\ .
\end{equation}
From this expression one can easily find
\begin{equation}\label{gvb}
\gamma=\sqrt{1+r^2 \chi^2}\hspace{0.3cm} {\rm and} \hspace{0.3cm}
v={r\chi\over\sqrt{1+r^2\chi^2}}\, .
\end{equation}

\begin{figure}[htb]
\begin{center}
\includegraphics[width=8cm]{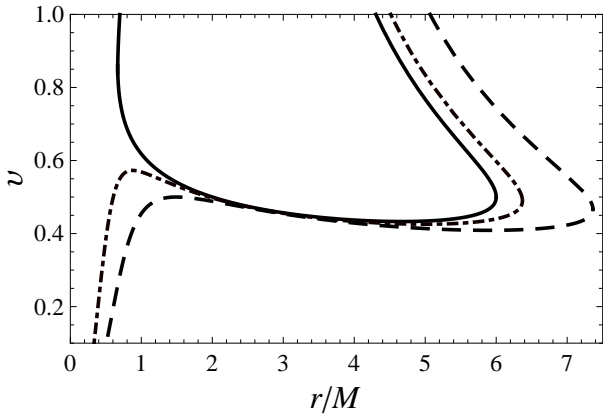}
\caption{\label{velocity} Velocity of the particles at $r_{_{\rm
ISCO}}$ as a function of its radius for different values of the
gravitomagnetic charge: $l/M=0$ (solid line), $l/M=0.4$
(dot-dashed line), $l/M=0.8$ (dashed line). }\label{fig_0}
\end{center}
\end{figure}

\begin{figure*}[htb]
\begin{center}
a) \includegraphics[width=8cm]{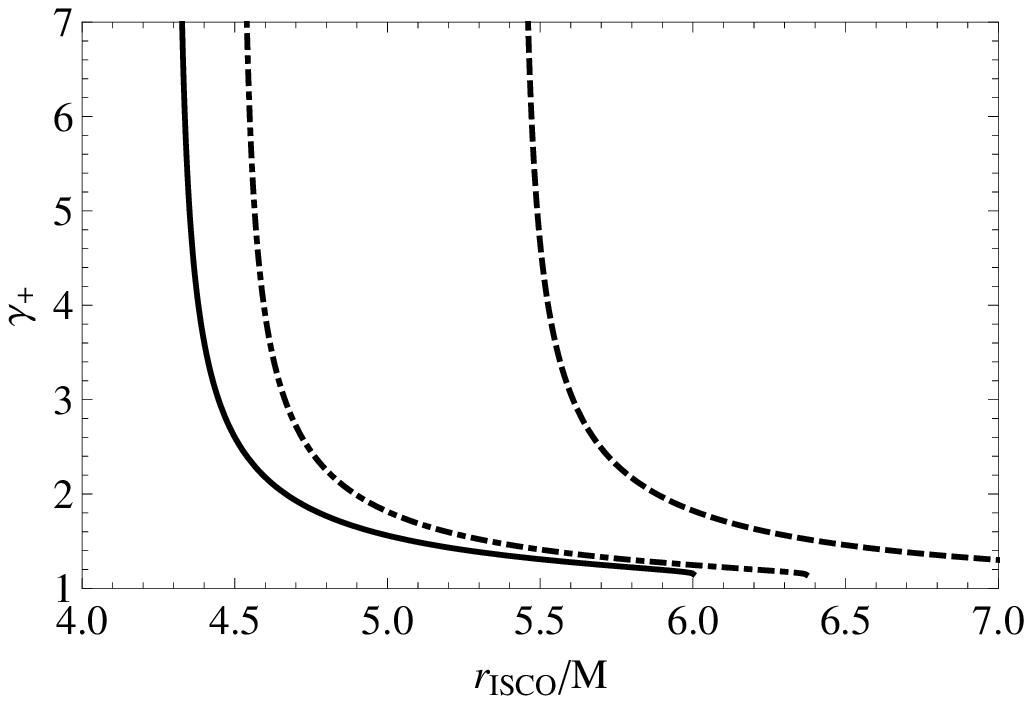} b)
\includegraphics[width=8cm]{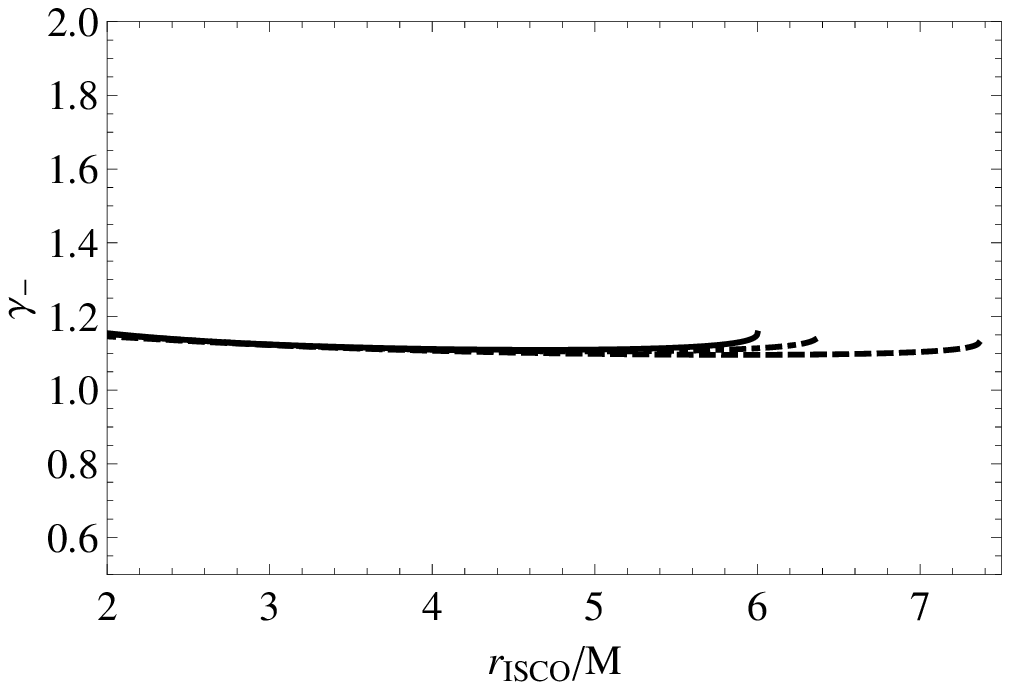}
\caption{\label{gammaplus}$\gamma_{+}$ (a) and $\gamma_{-}$ (b) as
a function of the ISCO radius for different values of the
gravitomagnetic charge: $l/M=0$ (solid line), $l/M=0.4$
(dot-dashed line), $l/M=0.8$ (dashed line). }\label{fig_0}
\end{center}
\end{figure*}

Using (\ref{gvb}) one can find the values of the velocity
$v_{\pm}$ and $\gamma$-factor $\gamma_{\pm}$. Fig.~\ref{velocity}
shows the velocity of a particle at the ISCO as a function of  its
radius, while Fig.~\ref{gammaplus} shows the dependence of
$\gamma_{\pm}$ from $r_{_{\rm ISCO}}$.

{The Fig.~\ref{velocity} shows the dependence of the velocity of
the particle at ISCO. Since there are two values of velocity for
the same radius one can interpret them as two values of velocity
which correspond for two opposite directions of motion of the
particles. Since charged particle and magnetic field interaction
depends on the velocity direction there should be two values
$v_{\pm}$ for each $r_{_{\rm ISCO}}$ . Furthermore  the absence of
external magnetic field (right border of the plots) one can obtain
only one solution for the velocity at ISCO. One should mention
that in the non-relativistic case one can get the Keplerian
velocity profile.}

As the next step, following to \citet{frolov} we will study the
center-of-mass collision of two particles in the vicinity of a
black hole with nonvanishing gravitomagnetic charge, when one of
these particles has the mass $m$ and charge $q$ and rotates along
the circular orbit. Another particle, which is neutral has a mass
$\mu$ and 4-momentum $\mathbf{k}$ and freely falls from the rest
at spatial infinity. From the conservation of momenta one can
write the momentum of the system at the
moment of
collision as
\begin{eqnarray}
\mathbf{P}=\mathbf{p}+\mathbf{k}\, .
\end{eqnarray}
This implyies that the center-of-mass
energy $E_{\rm c.m.}$ of two
colliding particles is
\begin{eqnarray}
E_{\rm c.m.}=m^2+\mu^2-2 (\mathbf{p},\mathbf{k})\, .
\end{eqnarray}

Using the equation of  particle motion around black hole with
nonvanishing gravitomagnetic charge (\ref{teq})--(\ref{req}) one
can obtain the following relation for the center-of-mass energy of
two colliding particles for the weak magnetic field approximation:
\begin{eqnarray}
\frac{E_{\rm c.m.}}{m}\simeq
0.3\sqrt{\frac{96-l^2}{\sqrt{8+l^2}}}b^{1/4}\ .
\end{eqnarray}

{In the Table~\ref{tab1} the dependence of the ISCO radius and the
center of mass energy of colliding charged particles have peen
shown. From the results on can conclude that gravitomagnetic
charge correction prevents the particle from the infinite
acceleration.}

\begin{table*}
\caption{\label{tab1} The dependence of the center of mass energy
and ISCO radii from the magnetic parameter $b$ for the different
values of the specific gravitomagnetic charge $l/M$:}

\begin{center}
{\begin{tabular}{@{}ccccccccc@{}} \hline\noalign{\smallskip}
 ${l/M}$ & 0 & 0.2 & 0.4 & 0.6 & 0.8 & 1.0 & 4.52 \\
\\
\noalign{\smallskip}\hline\noalign{\smallskip}
 ${ r_{_{\rm ISCO}}}$ & $2+0.58 b^{-1}$ & $2.02+0.58 b^{-1}$ & $2.08+0.59 b^{-1}$ & $2.17+0.60 b^{-1}$ & $2.28+0.61 b^{-1}$ & $2.41+0.62 b^{-1}$ & $5.63+0.74 b^{-1}$ \\
 \noalign{\smallskip}\hline\noalign{\smallskip}
 ${ E_{\rm c.m.}}/m$ & $1.747 b^{1/4}$ & $1.745 b^{1/4}$ &
 $1.738 b^{1/4}$ & $1.725 b^{1/4}$ & $1.708 b^{1/4}$ & $1.688 b^{1/4}$ & $1.129 b^{1/4}$   \\
 \noalign{\smallskip}
\end{tabular}}
\end{center}
\end{table*}

From the obtained result one can observe that the presence of the
gravitomagnetic charge will decrease the value of the center of
mass energy. The role of the magnetic
field in particle accelerating
process is to decrease the innermost
stable circular orbits radii.
As particles come closer
to the black hole
horizon their energy at
infinity is going to increase. The role of the gravitomagnetic
charge in this process is opposite: the presence of the
gravitomagnetic charge increase the radii of ISCO.

\section{Conclusion}
\label{conclusions}

In this report we have obtained the expressions for the energy and
angular momentum as well as ISCO of the charged particle in the
vicinity of the black hole  in presence of  gravitomagnetic charge
and exterior magnetic field.

Recently, \citet{Banados:2009pr} underlined
that a rotating black
hole can, in principle, accelerate
the particles falling to the
central black hole to arbitrary
high energies. \citet{frolov} has
shown that the magnetic field could play a role of charged
particle accelerator near the nonrotating black hole. Because of
some mechanisms such as astrophysical
limitations on the maximum
spin, backreaction
effects, upper limit for
magnetic field, and
sensitivity
to the initial conditions, there appears to
be some upper limit for the center of mass
energy of the infalling
particles. One of the mechanisms
offered in this paper is appearing due to the gravitomagnetic
charge correction which prevents the particle from the infinite
acceleration.

\acknowledgments

AAA and BJA thank the TIFR and IUCAA for the warm hospitality
during their stay in Mumbai and Pune, India. AAA and BJA
acknowledge the Akdeniz University  and TUBITAK/BIDEP foundation
for supporting local hospitality during their stay in Antalya,
Turkey. This research is supported in part by the projects
F2-FA-F113, FE2-FA-F134 of the UzAS and by the ICTP through the
OEA-PRJ-29 and BIPTUN projects. AAA and BJA acknowledge the German
Academic Exchange Service (DAAD) and the TWAS Associate grants.

%

%


%

\begin{thebibliography}


\bibitem[\protect\citeauthoryear{{Abdujabbarov et al.}}{2008}]{aak08}
Abdujabbarov, A.A., Ahmedov, B.J., Kagramanova, V.G.:
Gen. Rel. Grav. {\bf 40}, 2515 (2008)
%
%
\bibitem[\protect\citeauthoryear{{Abdujabbarov et al.}}{2011}]{shaymiev11}
{Abdujabbarov, A.A.,  Ahmedov, B.J., Shaymatov, S.R., Rakhmatov,
A.S.:  Astrophys Space Sci {\bf  334}, 237 (2011)}

%
\bibitem[\protect\citeauthoryear{{Ahmedov et. al}}{2012}]{aka12}{
Ahmedov, B.J., Khugaev, A.V., Abdujabbarov, A.A.: Astrophys Space
Sci {\bf 337}, 679 (2012) }
%
\bibitem[\protect\citeauthoryear{{Ahmedov and Rakhmatov}}{2003}]{ar03}
Ahmedov, B.J., Rakhmatov, N.I.: Found. Phys. \textbf{33}, 625
(2003)
%
\bibitem[\protect\citeauthoryear{{Aliev et al.}}{2008}]{aliev08}
{Aliev, A. N., Cebeci, H., Dereli, T.: {\prd} {\bf 77}, 124022
(2008)}
%
\bibitem[\protect\citeauthoryear{{Banados et al.}}{2009}]{Banados:2009pr}
{Banados, M., Silk, J., West, S.~M.: Phys.\ Rev.\ Lett.\  {\bf
103}, 111102 (2009)}
%
\bibitem[\protect\citeauthoryear{{Berti et al.}}{2009}]{berti}
{Berti, E., Cardoso, V., Gualtieri, L., Pretorius, F., Sperhake,
U.: {Phys. Rev. Lett.} \textbf{103}, 239001 (2009) }
%
\bibitem[\protect\citeauthoryear{Bini et al.}{2003}]{bini03}{Bini, D.,
Cherubini, C., Janzen, R.T., Mashhoon, B.: Class. Quantum Grav.
{\bf 20}, 457 (2003)}
%
\bibitem[\protect\citeauthoryear{Chowdhury et al.}{2012}]{joshi2}{
Chowdhury, A. N., Patil, M., Malafarina, D., Joshi, P. S.: Phys.
Rev. D, {\bf 85}, 104031 (2012)}
%
\bibitem[\protect\citeauthoryear{{Dadhich} and {Turakulov}}{2002}]{Dadhich02}
{Dadhich, N., Turakulov, Z.Ya.: {Class. Quantum Gravit.}
\textbf{20},457 (2003)}
%

%
\bibitem[\protect\citeauthoryear{Frolov}{2012}]{frolov}{
Frolov, V.P.: Phys. Rev. D {\bf 85}, 024020 (2012)}
%
\bibitem[\protect\citeauthoryear{Kagramanova et al.}{2010}]{kkhl10}{
Kagramanova, V., Kunz, J., Hackmann, E., Lammerzahl, C.: Phys.
Rev. D {\bf 81}, 124044 (2010)}
%
\bibitem[\protect\citeauthoryear{Kagramanova et al.}{2008}]{kkl08}
{Kagramanova, V., Kunz, J., L\"{a}mmerzahl, C.: {Class. Quantum
Grav. }{\bf 25}, 105023 (2008)}
%

\bibitem[\protect\citeauthoryear{Lynden-Bell and Nouri-Zonoz}{1998}]{lbnz98}{Lynden-Bell, D., Nouri-Zonoz,
M.: Rev. Mod. Phys., {\bf 70}, 427 (1998)}
%
\bibitem[\protect\citeauthoryear{Morozova et al.}{2008}]{mak08}
{Morozova, V. S., Ahmedov, B. J. and  Kagramanova, V. G.:
{Astrophys. J.} {\bf 684}, 1359 (2008).}
%
%
\bibitem[\protect\citeauthoryear{{Morozova} and {Ahmedov}}{2009}]{ma08}
%
{Morozova, V.S., Ahmedov, B.J.: {Int. J. Mod. Phys. D} {\bf 18},
107 (2009)}
%

\bibitem[\protect\citeauthoryear{{Newman et al.}}{1963}]{nut63}
Newman, E., Tamburino, L., Unti, T.: J. Math. Phys. \textbf{4},
915 (1963)


%
%
\bibitem[\protect\citeauthoryear{Nouri-Zonoz}{2004}]{zonoz07}
{Nouri-Zonoz, M.:  {Class. Quantum Grav.} \textbf{21}, 471 (2004)}
%
%

\bibitem[\protect\citeauthoryear{Nouri-Zonoz et. al}{1999}]{nzdlb99}{
Nouri-Zonoz, M., Dadhich, N., Lynden-Bell, D.: Class. Quantum
Grav., {\bf 16}, 1021 (1999)}
%

\bibitem[\protect\citeauthoryear{Nouri-Zonoz and Lynden-Bell}{1997}]{nzlb97}{
Nouri-Zonoz, M., Lynden-Bell, D.: Mon. Not. R. Astron. Soc., {\bf
292}, p. 714}
%
%
\bibitem[\protect\citeauthoryear{Patil and Joshi}{2012}]{joshi1}{
Patil, M., Joshi, P. S.: Phys. Rev. D, {\bf 85}, 104014 (2012)}
%

\bibitem[\protect\citeauthoryear{Penrose}{1969}]{pnr74}
{Penrose, R.: Rivista del Nuovo Cimento {\bf 1}, 252 (1969)}
%
\bibitem[\protect\citeauthoryear{Piran et. al}{1975}]{Piran:1975}
{Piran, T.,  Katz, J. and Shaham, J.:  Astrophys. J. Lett., {\bf
196}, L107 (1975)}
%
%
\bibitem[\protect\citeauthoryear{Piran and Shaham}{1977a}]{Piran:1977dm}{Piran, T.,
Shaham, J.: Phys.\ Rev.\  D {\bf 16}, 1615 (1977a)}

\bibitem[\protect\citeauthoryear{Piran and Shaham}{1977b}]{Piran:1977aa}{
Piran, T., Shaham, J. Astrophys. J., {\bf 214}, 268 (1977)}
%

\bibitem[\protect\citeauthoryear{Piotrovich et.al}{2011}]{psgn11}{
Piotrovich, M. Yu., Silantev, N. A., Gnedin, Yu. N.,
Natsvlishvili, T. M.: Astrophys. Bulletin, {\bf 66}, 320 (2011)}
%
\bibitem[\protect\citeauthoryear{{Rahvar} and {Habibi}}{2004}]{habibi04}
{Rahvar, S., Habibi, F.: Astroph. J., {\bf 610}, 673 (2004)}
%
\bibitem[\protect\citeauthoryear{{Zimmerman} and {Shahir}}{1989}]{Zim89}
{Zimmerman, R.L., Shahir, B.Y.: {Gen. Relativ. Gravit.}
\textbf{21},821 (1989).}
%

\end{thebibliography}

%

\end{document}